\documentclass{elsart}
\usepackage{graphicx}
\newcommand{\epem}{e$^{+}$e$^{-}$}
\begin{document}
\begin{frontmatter}
\title{Dark photon searches using displaced vertices at low energy 
\epem\ colliders}
\author{Fabio Bossi}
\address{Laboratori Nazionali dell'INFN Frascati}
\begin{abstract}
The existence of a new, photon-like, massive particle, 
the $\gamma'$ or dark photon,
is postulated in several extensions of the Standard Model. These models 
are often advocated to
explain some recent puzzling astrophysical observations, as well as
to solve the unsofar unexplained deviation between the measured and 
calculated values of the muon anomaly. 
Dark photons can be produced at \epem\ colliders both in continuum
events and in vector meson transitions and can eventually decay into
an electron-positron pair.    
For a proper choice of the parameters of the theory, 
a $\gamma'$ can have a relatively long lifetime and can therefore 
be observed as an \epem\ vertex well separated by the primary interaction 
point. This case is discussed in reference to very high luminosity
\epem\ colliders either in construction or under study in several 
laboratories in the world. It is shown that a search strategy based 
on the detection of displaced vertices can be in principle very effective 
in covering a rather wide and to date unexplored region of the theoretical 
parameters space
\end{abstract}     
\end{frontmatter}

\section{Introduction}
In the Standard Model (SM), interactions among elementary particles 
are mediated 
by the vector bosons of the strong, weak and electomagnetic forces. 
Experimental evidence for the existence of those bosons is compelling 
and  precise measurements of their properties have been accumulated in the 
past decades.
New forces can have escaped detection so far, either if their 
associated bosons are very heavy or if their couplings to ordinary matter
are weak enough. The latter case has been advocated, among others,
 in models which try to 
explain and reconcile among them several puzzling astrophysical observations
performed in recent years~\cite{Ark,Pos1,Cel,Bor}. They are sometimes also 
used to reconcile the measured value of the muon anomaly to the 
SM prediction, which differ by approximately by 
3.5 $\sigma$ (see, for instance~\cite{Pos4}). 

If new, light, neutral bosons (which from now on will be called $\gamma'$
or {\it dark photons})
exist and if they are measurably, albeit weakly,
coupled with SM particles they can be produced and observed at colliding-beams
and fixed targed experiments~\cite{Fay1,Best,Pos2,RWa,Pos3,Est,Nic}. 
In fact there have been several attempts to observe evidence 
for such particles, using data from running 
facilities~\cite{A1,Apx,BaB1,BaB2,Kl1,Kl2,Was} or data-mining old
experiments~\cite{Best,Sar,Blu,Gn1,Gn2}. 
Since no evidence for their existence was found, limits have been set as a 
function of the $\gamma'$ mass and of its coupling 
to ordinary matter. 

In the near future, new experiments under construction 
 are expected to extend those limits in a 
region of couplings and/or masses so far unexplored. 
All of them are designed to exploit the 
radiative production of the $\gamma'$ by a very intense electron or 
positron beam on 
a properly built high-Z target~\cite{Ap1,HPS,MESA,Dlg, VEPP3,gnc}.      
The purpose of the present letter is to show that comparable results can 
be obtained by high luminosity, low energy electron-positron colliders, such as
those under construction or under study in several laboratories in the
world~\cite{df2,itc,spbl}.
 These facilities will take advantage of two main construction
features which coherently conspire to enhance their discovery potential:
their very high goal luminosity, and the usage of very compact beams 
(these two features are in fact strongly correlated).  
Actually, high luminosity translates into the possibility of probing lower 
production 
cross sections, i.e. lower effective couplings between the 
$\gamma'$ and ordinary matter. On the other hand, low couplings  
translate into longer $\gamma'$ decay paths, 
especially for low $\gamma'$ masses. 
Thus, the usage of beams of very small dimensions 
allows one to obtain a clear $\gamma'$ signal  by observing  
secondary vertices of a well defined invariant mass,  
well separated by the beams interaction point.

In the paper,  this case will be discussed for 
three different possible choices of the machine center-of-mass energy, 
corresponding, respectively, to the mass of the $\phi$(1020),
the J$/\psi$(1S) and the $\Upsilon$(4S) mesons. 
This choice is motivated by the projects mentioned above.
It will be shown that higher energy machines are favoured, not only 
because they are expected to deliver larger data sets, but also because
the $\gamma'$ therein produced have longer decay paths, ceteris paribus.
Instrumental effects play however a relevant role in the actual
detection strategy, and can in some cases dramatically reduce
the discovery potential of the method. Still, in particular for the
case of a high luminosity $\tau$-charm factory, it remains high enough to
be competitive to the fixed target experiments mentioned above. 

The paper is organized as follows. Firstly, the theoretical framework 
of the paper is discussed, together with a 
short presentation of the experimental limits on the existence of 
dark photons obtained so far. 
The case for the searches at low energy, high luminosity \epem\ colliders 
is discussed in section 3, followed by some considerations on the actual
implementation of the proposed method to existing, or planned, facilities.
Radiative vector meson decays are discussed in section 5. Conclusions 
are given in section 6.    

\section{Physics case}
In many new physics scenarios, the SM is extended by simply adding an 
additional U(1)$_{D}$ symmetry, under which SM particles are uncharged at 
first order~\cite{Pos4,Fay2,Hold}. 
The gauge boson associated to the new symmetry, 
the $\gamma'$, can still interact with ordinary 
matter via kinetic mixing described by an effective interaction Lagrangian

\begin{equation}
L_{kin-mix} = i\epsilon~e~\overline{\psi}_{SM}~\gamma^{\mu}~\psi_{SM}~A_{\mu} 
\label{eq:kinmix}
\end{equation} 

where $A$ denotes the $\gamma'$ field. 
The kinetic mixing factor $\epsilon$ parametrizes the coupling strength 
relative to the electric charge and is predicted in various models to
be in the range 10$^{-12}$ - 10$^{-2}$. The mass of the dark photon 
(m$_{\gamma'}$) rests unpredicted.  On phenomenological grounds, however, 
 masses in the MeV-GeV range,
which are of interest for the present work, are favoured. 

There might exist non-SM matter particles which are sensitive to the 
new U(1)$_{D}$ interaction. Often they are postulated to be the main 
constituent of the yet undiscovered Dark Matter component of the Universe (DM), 
and must therefore be electrically neutral and stable. 
If kinematically allowed, the $\gamma'$ will decay preferably into pairs of 
these particles, thus its decay becomes ``invisible''.
The case for detecting invisible decays is discussed, for instance, 
in~\cite{VEPP3, Dl2, dnv, npdm, Ess}. 
 On the other hand, 
if the dark photon is lighter than DM, it is forced to decay into
a pair of SM particles, with a width regulated by eq.~\ref{eq:kinmix}.   
In this case, its proper time is approximately given by~\cite{Sarah}

\begin{equation}
c\tau (mm) = \frac{0.08}{N_{f}}\cdot(\frac{10^{-4}}{\epsilon})^{2}
\cdot\frac{100}{m_{\gamma'}(MeV)} 
\label{eq:lifet}
\end{equation}

where N$_{f}$ is the number of SM decay channels allowed by kinematics.

\begin{figure}[h]
%\begin{centering}
\includegraphics[width=.8\textwidth]{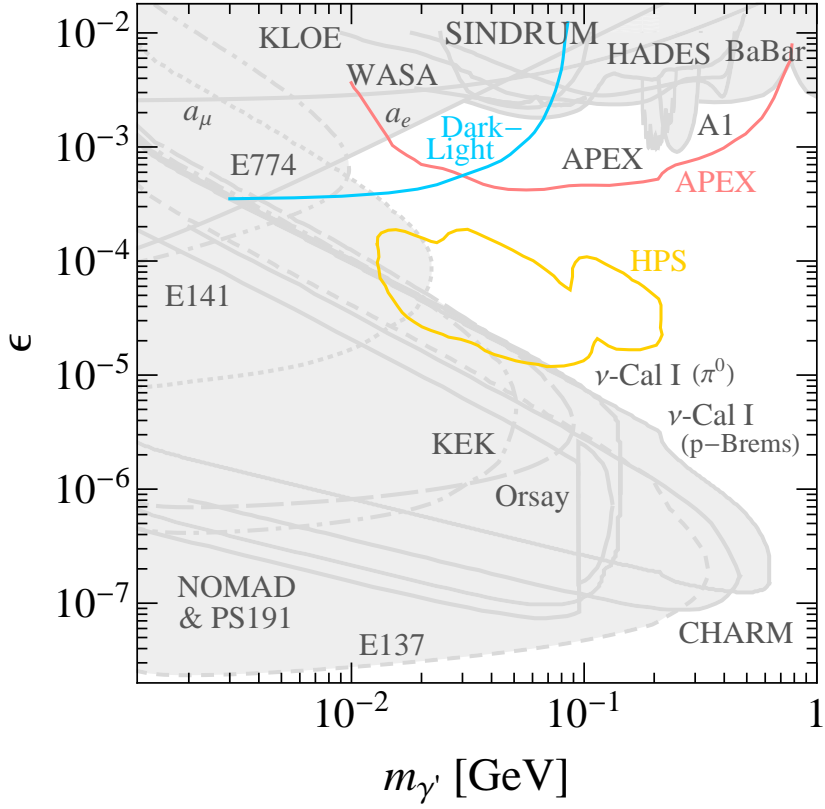}
\caption{Excluded region in the plane $\epsilon-m_{\gamma'}$ resulting
from presently available data. 
Electron beam dump experiment cover the region of low masses and very low
couplings, down to $\epsilon \sim$10$^{-7}$. For higher masses and lower 
couplings limits come mainly from meson decays, electron-nucleon 
scattering experiments and to B-factories data
 (plot courtesy of S. Andreas). The plot reports also the projections
for the experiments presently under construction. 
For details on single experiments see \cite{Sarah}.}
\label{fig:sarah}
\end{figure}

There have been several attempts to experimentally observe a $\gamma'$ 
signal, using many different techniques.  
Figure~\ref{fig:sarah}, taken from reference~\cite{Sarah}, 
shows the exclusion plot in the plane 
$m_{\gamma'}$-$\epsilon$, resulting from the above mentioned searches. 
Electron beam dump experiment cover the region of low masses and very low
couplings, down to $\epsilon \sim$10$^{-7}$. For higher masses and lower 
couplings limits come mainly from meson decays, electron-nucleon 
scattering experiments and to B-factories data.
Important information, not shown in figure~\ref{fig:sarah}, 
 can also be deduced by astrophysical observations
(see for instance~\cite{Ast1} and references therein). 
For m$_{\gamma'}>$10-20 MeV, the region with
 $\epsilon<$10$^{-3}$ remains largely unexplored. 

There are currently various 
experiments either running or under construction aiming at probing part 
of this region. All of them are designed to
 exploit the 
radiative production of the $\gamma'$ by a very intense electron or
positron beam on 
a properly built high-Z target.      
 In particular the HPS experiment at Thomas Jefferson Laboratory
(USA) is designed to cover the region $\epsilon$=(10$^{-4}$,10$^{-5}$), 
m$_{\gamma'}$=(20,300) MeV~\cite{HPS} (see figure~\ref{fig:sarah}).
 
In the following, the case for the search 
for dark photons in the same parameters space region at a very high 
luminosity \epem\ collider will be discussed.  
      
\section{Searches at \epem\ colliders}
In the last decades large amount of data have been collected
at high luminosity \epem\ 
flavour factories operating  at different 
center of mass energies.
These data range from the $\sim$2 fb$^{-1}$ delivered at the $\phi$(1020) 
peak by the italian collider DA$\Phi$NE, to the 0.5-1 ab$^{-1}$ produced 
by the B-factories at PEP-II (USA) and KEK-B (Japan).  In the near future a
consistent increase of the above statistics is expected both at 
 DA$\Phi$NE and at KEK-B which aim at increasing their data sample by a 
factor of 10 and 50, respectively. An option to increase the center of mass 
energy of DA$\Phi$NE up to 2.5 GeV, has been taken into 
consideration~\cite{d25}.   
Finally, studies for the construction of
a collider capable of delivering $\sim$ 1 ab$^{-1}$  around the charm threshold
are under consideration in Italy, Russia and China 
(see for instance~\cite{itc}). 

As of today, searches for dark photons at \epem\ colliders have been
pursued mainly  by studying the process \epem $\rightarrow \gamma\gamma'$ with
the subsequent decay of the $\gamma'$ into a $\mu^{+}\mu^{-}$ pair.
This limits the search to m$_{\gamma'}>$2~m$_{\mu}$, which, as a consequence
of equation~\ref{eq:lifet}, results in its lifetime being unmeasurably short. 
Therefore the signal can be separated by the more copious and 
otherwise indistinguishible  QED background, only by observing a sharp peak 
 in the invariant mass distribution of the final state 
lepton pair. 

The question arises whether it would be possible to extend the search also to 
the region with m$_{\gamma'}<$2~m$_{\mu}$ and in particular with 
$\epsilon<$10$^{-3}$.    
The main message of the present paper is that the foreseen increase of the 
potentially available data sample allows one to give a positive answer to 
the above question,   
not only because of the increased statistical sensitivity but also because
it opens the doors to the possibility of observing a clear signal for a
long-lived $\gamma'$, which is only marginal with the presently 
available data. 

Here and in the following, for the sake of simplicity, 
only the case of symmetric machines is considered.
Also, since we are interested at the case with m$_{\gamma'}<$  2m$_{\mu}$, 
the dark photon can decay only into a \epem\ pair.
   
The differential cross section for radiative $\gamma'$ production in
\epem\ collisions, is given by~\cite{Est} 

\begin{equation}
\frac{d\sigma}{dcos~\theta}=\frac{2~\pi~\alpha^{2}~\epsilon^{2}}{E_{c.m.}}
\cdot \frac{1 + cos^{2}~\theta}{sin^{2}~\theta}
\label{eq:xsect}
\end{equation}

where $\theta$ is the angle between the incoming positron and the outcoming
photon and E$_{c.m.}$ denotes the center of mass energy of the event. 
By integrating the above equation between cos($\theta$)= $\pm$ 0.8, 
one obtains a total cross section of approximately 
0.9$\times \epsilon^{2}$ $\mu$b, 
0.1$\times \epsilon^{2}$ $\mu$b and
3.6$\times \epsilon^{2}$ nb
for E$_{c.m.}$ = 1.02, 3.1, 10.5 GeV respectively. 
Thus, in the presently 
available data samples, for $\epsilon$=10$^{-4}$ a few dozens $\gamma'$ 
can be found.

Due to the two body kinematics, the dark photon is boosted in the laboratory
frame by a factor E$_{c.m.}$/m$_{\gamma'}$.  Therefore for small  
kinetic mixings  and for low enough dark photon 
masses, its lifetime in the laboratory frame becomes sizeable.    
For instance, for $\epsilon$=10$^{-4}$ and m$_{\gamma'}$=20 MeV, the mean decay
path of a dark photon is $\sim$1, 3, 10 cm  for E$_{c.m.}$ = 1.02, 3.1, 10.5 
GeV respectively. 

Can these long decay paths be exploited to separate a potential $\gamma'$ 
signal from the QED background? Clearly, although the {\em secondary}
vertex can be determined with standard vertexing techniques, the {\em
primary production} one cannot on an event by event basis.   
On the other hand, the actual position and size of the collision envelope 
can be 
determined {\em on a statistical basis} using known processes, such as Bhabha 
scattering or muon pair production. 
Interestingly, at all of the \epem\ facilities
under consideration, one of the strategies used to maximize luminosity is to
keep the transverse beam dimensions
at the interaction point as small as possible, typically $\leq$1 mm. 
 Therefore, 
assuming a perfectly gaussian distribution of the beam spot, with 
a maximum transverse dimension of 1 mm, the probability for observing 
an \epem\ vertex from standard QED processes at a transverse distance 
of 1 cm or more from the center of the collision 
spot is practically zero. On the other hand, 
the number of $\gamma'$ decay events with transverse decay path larger 
than 1 cm, N$_{1cm}$,  can be as large as several thousands, 
depending on the actual value
of $\epsilon$, m$_{\gamma'}$, E$_{c.m.}$ and of the luminosity integrated 
by the machine, $L_{int}$. 

\begin{figure}[h]
%\begin{centering}
\includegraphics[width=.8\textwidth]{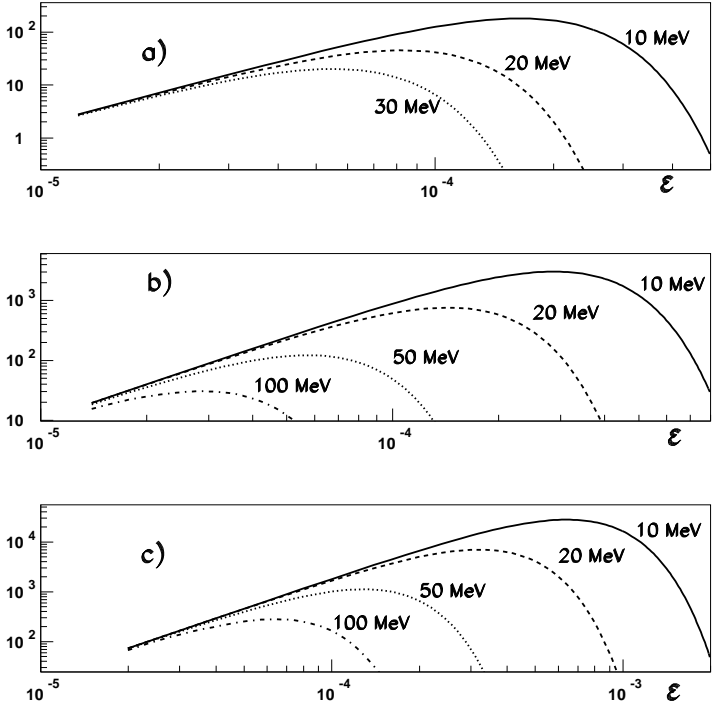}
\caption{Number of $\gamma'$ decay events with transverse decay path larger 
than 1 cm, as a function of $\epsilon$, for different values of m$_{\gamma'}$. 
Three different experimental conditions are taken into consideration:
a) E$_{c.m.}$ = 1.02 GeV, L$_{int}$=20 fb$^{-1}$; 
b) E$_{c.m.}$ = 3.1 GeV, L$_{int}$=1 ab$^{-1}$;
c) E$_{c.m.}$ = 10.5 GeV, L$_{int}$=50 ab$^{-1}$. Note that the horizontal
scale is slightly different for the three plots.}
\label{fig:fig2}
\end{figure}

Figure~\ref{fig:fig2} shows 
the variation of N$_{1cm}$ as a function of $\epsilon$, for different 
values of m$_{\gamma'}$ and for three different experimental conditions:
a) E$_{c.m.}$ = 1.02 GeV, L$_{int}$=20 fb$^{-1}$; 
b) E$_{c.m.}$ = 3.1 GeV, L$_{int}$=1 ab$^{-1}$;
c) E$_{c.m.}$ = 10.5 GeV, L$_{int}$=50 ab$^{-1}$.
The choosen values for L$_{int}$ correspond to the target performance for
the facilities under construction or under study mentioned above. 
The behaviour of the curves is easily explained. For $\epsilon \ll$10$^{-4}$
the mean decay path of a dark photon is much larger than 1 cm, 
and therefore N$_{1cm}$ 
increases with $\epsilon^{2}$,  independently of $m_{\gamma'}$. It eventually 
reaches a peak and drops rapidly towards zero, as long as, with increasing 
$\epsilon$, the lifetime becomes shorter and shorter. The position of the 
peak is determined by the proper balance between the effect of the 
production cross section, which increases with $\epsilon^{2}$, and that of the
lifetime which decreases with it. It depends also on the value of 
m$_{\gamma'}$, the decay path decreasing again quadratically with it.        
It is seen that, despite the lower production cross section, 
the largest expected  integrated luminosity combined  with the higher 
boost factors, favour the B-factory option (case c)). In this case, 
however, the peak
of the distribution, especially for lower masses, is obtained for 
values of the kinetic mixing $\sim$10$^{-3}$. It can be noted also that 
in case a) the number of observable  dark photons with masses greater than
$\sim$ 30 MeV becomes hopelessly small. This is not only due to the lower
luminosity but also to the reduced Lorentz boost, consequence of
 the lower center of mass energy of the collision. 

Although the results obtained so far look very encouraging on general 
grounds, there are two main limitations coming from the implementation
of the above search strategy into a real experiment. 
On the one hand, for specific values of the parameters, the $\gamma'$ 
lifetime 
becomes so long that a relevant part of the decays would escape detection
of an apparatus of realistic dimensions. For instance, 
for E$_{c.m.}$= 10.5 GeV, $\epsilon$=5$\times$10$^{-5}$ and 
m$_{\gamma'}$=10 MeV, the $\gamma'$ mean decay path is about 1.5 m. 
More importantly, a very dangerous {\em instrumental} background 
comes into operation, namely photon conversions on the detector 
material induced by \epem $\rightarrow \gamma\gamma$ events. This is 
particularly relevant since experiments are often designed to have beam pipes 
with very small radii at the interaction point. Although one can try to
minimize the conversion probability on the detector elements, 
by properly choosing type and dimensions of the materials, the cross 
section of the  \epem $\rightarrow \gamma\gamma$ process is so much larger
than the signal one (in fact it is larger by a factor 
$\epsilon^{-2}$) that this background rapidly  becomes unbeatable.  

\begin{figure}[h]
%\begin{centering}
\includegraphics[width=.8\textwidth]{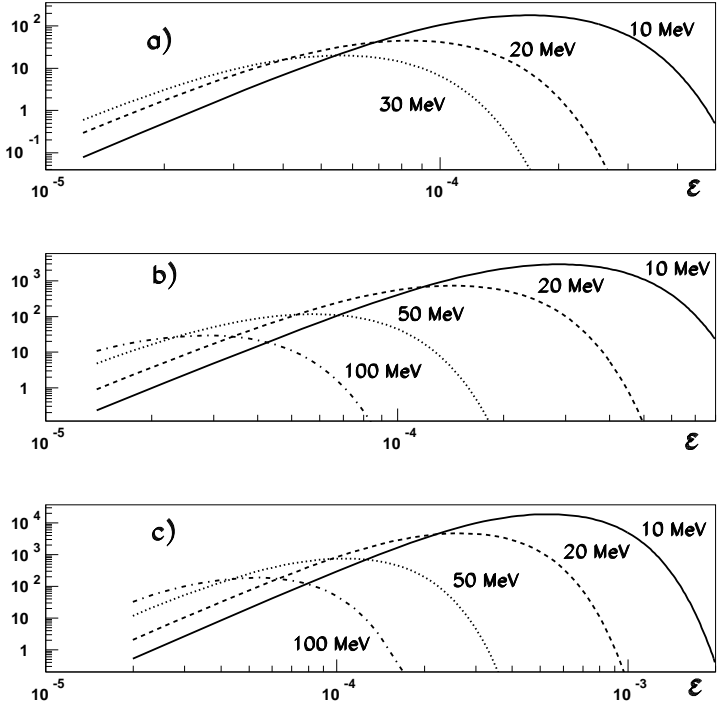}
\caption{Same as fig.~\ref{fig:fig2}, with the further request that the 
transverse decay path is lower than 5 cm.}
\label{fig:fig3}
\end{figure}

The simplest way to cope with this  problem is to allow for a large enough 
empty region around the interaction point, where photons cannot interact 
with matter and dark photons can at least partly undergo their decay.  
 It would then be reasonable to  accept 
only events with decay vertices occuring before the beam pipe but still far
(1 cm) from the nominal beam spot center. Assuming a beam pipe 
of 5 cm radius, such as that presently used by the KLOE-2 experiment 
at DA$\Phi$NE, the number of events thus obtained within acceptance, 
n$_{acc}$, 
is shown in figure~\ref{fig:fig3} for the three cases under consideration. 

For high values of $\epsilon$ this acceptance cut does not observably 
affect the previous distributions. In fact, in this case, the lifetime is 
so short that almost all of the dark photons decay much before 5 cm. 
For lower values of $\epsilon$, instead, the consequence of the cut in 
acceptance are more visible and can decrease the number of accepted
events by an order of magnitude, especially for very low $\gamma'$ masses. 
However, and this is one the main messages of this paper, the number 
of potentially observable events remains still considerable for a wide
region of the parameter space, especially for the higher energy machine 
options.  In particular, also allowing for some further detection 
inefficiency, it can be seen that kinetic mixings down to few times
10$^{-5}$ and masses up to $\sim$ 200 MeV can be probed. 

\begin{figure}[h]
%\begin{centering}
\includegraphics[width=.8\textwidth]{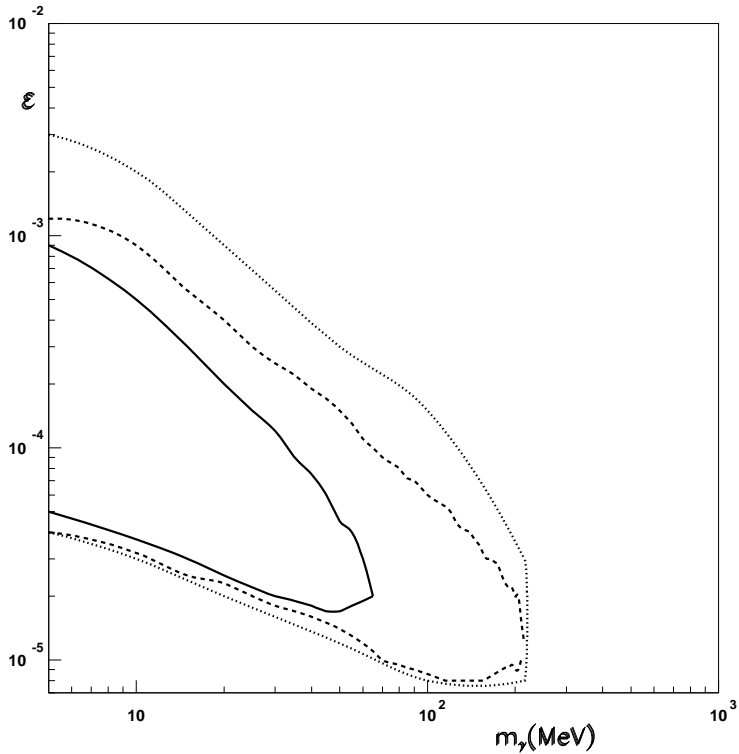}
\caption{Explorable region for cases a) (solid) b) (dashed) and c) (dotted) in
the plane $\epsilon$-m$_{\gamma'}$. No instrumental backgrounds are taken
into consideration, as well as potential efficiency and resolution effects
for the detector. Above $\sim$210 MeV, the opening of the 2-muon channel
drastically reduces the effectiveness of the method.}
\label{fig:fig5}
\end{figure}

Based only on signal statistics (i.e. without taking into consideration
possible detector resolution effects and other possible instrumental
backgrounds), this translates into the explorable regions shown
in figure~\ref{fig:fig5}, for the three cases under consideration. 
While case a) covers almost entirely a region already excluded
by previous beam-dump experiments, cases b) and c) can potentially
probe a relatively wide unexplored region (cfr. figure~\ref{fig:sarah}).
On the other hand, it has also to be underlined that this same region 
is expected to be covered by the aforementioned future fixed target 
experiments (see again figure~\ref{fig:sarah}). 

It is worthwhile stressing once more that the requirement for observing
a cm-scale decay path, ideally rejects every possible {\em physical}
background to our signal. Still, other instrumental effects need to be
taken carefully into account, as discussed in the following section.  

\section{Implementation at current and future facilities}
It is of interest understanding how difficult would it be to
practically implement on real experimental facilities  the 
ideas discussed so far. This requires a detailed knowledge of the actual 
machine and detector's design and expected (or measured) performance. Only
specific studies based on these figures can in the end determine whether 
the method is applicable or not, to which extent and on which machine.   
An obvious difference between our simplified models and reality can be found 
for instance in case c); both the old and the future B-factories 
are in fact asymmetric machines, the electron
beam being of higher energy with respect to the positron one. 
Although this might somewhat change the specific acceptance requirements 
with respect to the symmetric option discussed in this paper, 
it is however reasonable 
to assume that similar conclusions can still be drawn.

On general grounds, there are four parameters that have to be taken
into consideration: the primary $\gamma'$ production rate, the dimensions
of the beams, those of the beam pipe, and the vertexing capabilities
of the detector. 

As for the first point, 
despite the higher production cross section, the $\phi$-factory option 
(case a)) is
less performing than the other two, not only because of the much 
lower integrated luminosity, but also, as noted before, because of the 
intrinsic limitation due to the lower Lorentz boost factors. The proposal
for running DA$\Phi$NE at higher energies is, under this respect, particularly
interesting because this would allow increasing the decay paths proportionally
to E$_{c.m.}$.
 
As previously remarked, for all of the machines under consideration, 
the dimensions of the beams 
are kept very small in the transverse direction. For instance, at
 DA$\Phi$NE the beams have
$\sigma_{X}~\sim$1.5 mm, $\sigma_{y}~\sim$0.02 mm, and much lower dimensions
at the other machines.  
Note that both fig~\ref{fig:fig2} and fig~\ref{fig:fig3}, on which
we base our search strategy, refer to  
transverse decay paths, so the beam dimensions in the longitudinal 
direction are irrelevant for our conclusions. 
Obviously, non gaussian tails of the collision envelope can to some
 extent increase 
the background contamination. However, if not completely suppressed, they can
still be studied using other known processes, as for the gaussian part.

A real concern are the actual beam-pipe dimensions. 
Among the existing facilities, KLOE-2 at DA$\Phi$NE is the only
one having a beam pipe radius at the interaction point of 5 cm.
As for the machines running at the charm threshold, the only one presently
in operation, the chinese collider BEPC, has a beam pipe at the interaction
point of 3.5 cm in radius; however its present luminosity is about a 
factor 100 lower than required by our arguments at that energy. 
For the B-factories, the beam pipe radii range 
from the 2.5 cm for the BaBar detector to 1 cm for the future 
SuperBelle.

While a reconsideration of the inner region of SuperBelle is
probably very unlikely, in the case of a 
future high luminosity $\tau$-charm factory  it is conceivable that  
the interaction region can be designed such as to maximise the 
sensitivity for the dark photon search under consideration. 
It is worth noticing here that the request for minimizing the beam pipe
dimensions comes to first order from the experiment more than from the 
machine. In fact, they are somehow anti-correlated with the 
detector's vertexing capabilities. Actually, the resolution of 
a decay length measurement for a generic detector is approximately
proportional to the single point resolution of the most internal tracking 
device and inversely proportional to its distance from the decay 
point. Under this respect the less favourable situation is that of
KLOE-2, whose first tracking device, a triple-GEM cylindrical detector, has
an internal radius of 12 cm and single point resolution of $\sim$200$\mu$m. 
Still, its estimated vertex resolution for $K^{0} \rightarrow \pi^{+}\pi^{-}$
events is of 1-2 mm~\cite{It2}. 
The use of silicon detectors, which can have single 
point resolution of order 10 $\mu$m, would definitely improve with respect to
the KLOE-2 case. Noticeably, all the LEP experiments, which had beam pipes of 
5.5 cm, could reach typical decay length resolution of $\sim$250$\mu$m for 
B decay events, thanks to the use of silicon detectors~\cite{cosc}.
This implies that, considering the cm scale decay length we have been 
interested so far, vertexing resolution should not be a major issue. 
On the other hand, it can play a relevant role in the considerations
discussed in the following section. 
    
\section{Meson decays}
Electron-positron colliders provide a useful $\gamma'$ production mechanism
via radiative vector meson decays, too. Actually, for each observed 
$V \rightarrow P\gamma$ decay ($V$ and $P$ being a vector and a pseudoscalar
meson, respectively), there could be a $V \rightarrow P\gamma'$ process, 
suppresed by a factor $\epsilon^{2}$ with respect to the 
former one~\cite{RWa}. 
This fact has actually been exploited by the KLOE-2 Collaboration  which has 
searched for the dark photon using the $\phi \rightarrow \eta$~\epem\ process 
in~\cite{Kl1,Kl2}. As for the searches in the 
\epem $\rightarrow \mu^{+}\mu^{-}\gamma$
channel, the signal is separated by the SM (Dalitz decay) background
by looking for a peak in the invariant mass distribution of the final state
lepton pair.
However, once the $\gamma'$ lifetime becomes sizeable, 
these events are characterised as well by the presence of \epem\ vertices 
clearly separated by the collision point, thus allowing the usage of 
the search strategy described in the previous sections. 

The number of produced $\gamma'$ is given by 
\begin{equation}
N_{\gamma'} = N_{V}\cdot~BR_{VP\gamma}~\cdot~\epsilon^{2}
\end{equation} 

where N$_{V}$ is the number of produced vector mesons, and 
BR$_{VP\gamma}$ is 
the branching ratio for the corresponding standard radiative decay.
 
\begin{figure}[htp]
%\begin{centering}
\includegraphics[width=.8\textwidth]{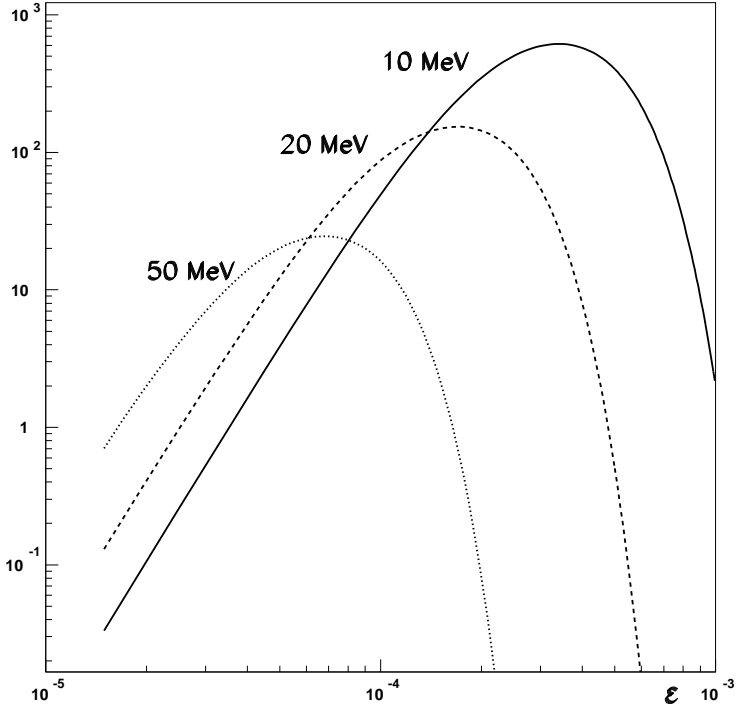}
\caption{Number of dark photons from the process
$J/\psi \rightarrow  \eta'\gamma'$
with decay paths larger than 1 cm and lower than 5 cm, as a function of
$\epsilon$ and for various values of m$_{\gamma'}$. 
An integrated luminosity of 1 ab$^{-1}$ is considered. }
\label{fig:fig4}
\end{figure}

Let us firstly consider the above mentioned 
$\phi \rightarrow \eta\gamma'$ process.
At a $\phi$-factory, $\sim$3$\cdot$10$^{9}$ $\phi$ mesons are
produced every fb$^{-1}$ delivered by the machine.  
Since BR$_{\phi\rightarrow\eta\gamma}~\simeq$1.3$\cdot$10$^{-2}$~\cite{PDG}, 
it is easily  seen that the number of produced signal
events, assuming  L$_{int}$=20 fb$^{-1}$, becomes negligible for 
$\epsilon \leq$10$^{-4}$. On the other hand, for higher values of 
$\epsilon$, the $\gamma'$  mean decay path becomes unmeasurably (as compared
with millimiter scale vertex resolutions) short, but for very low masses.
For instance, for $\epsilon$=2$\cdot$10$^{-4}$ it is already 0.8(0.2) cm, for
m$_{\gamma'}$=10(20) MeV. Unless, therefore, one integrates luminosities 
largely exceeding those expected from the presently considered machine, the 
method is hardly applicable to this decay channel.     

 Let us now turn our attention to the 
$J/\psi \rightarrow  \eta'\gamma'$ transition.
This process has been already studied  
in~\cite{FuLi}, where, however, the case for short lived 
dark photons only is considered.  
As before, one has
 N$_{J/\psi}~\simeq$~3$\cdot$10$^{9}$/fb$^{-1}$, if running at the J/$\psi$ 
peak.
 Considering that 
 BR$_{J/\psi \rightarrow \eta'~\gamma}~\simeq$5$\cdot$10$^{-3}$~\cite{PDG}, 
 one obtains $\sim$150 events for $\epsilon$=10$^{-4}$ and 
 L$_{int}$=1 ab$^{-1}$.  On the other hand, for this value of kinetic 
mixing, the $\gamma'$ mean decay path is of the right order of magnitude
only for a limited range of mass values. It is, for instance, 11.2, 2.8, 0.45 
cm for m$_{\gamma'}$=10, 20, 50 MeV respectively. The effect of this is 
seen in figure~\ref{fig:fig4} where the number of $\gamma'$ decays 
occurring at a
distance  between 1 and 5 cm from the interaction point is plotted as a 
function of $\epsilon$, 
for different values of m$_{\gamma'}$. Differently from the \epem$\gamma$
case, here it is the effect of the lifetime being short to  be dominant, 
at least for the kinetic mixing values of interest,  because  
for higher dark photon masses and low enough $\epsilon$, almost all of 
the $\gamma'$ survive for less than 1 cm. 
Note, also, that some further reduction in the number of observable 
events must be considered due to geometrical acceptance considerations. 
Still, there remain
a small region of the parameter space for which one can hope to observe 
a reasonable number of $\gamma'$ decays within acceptance. 

There is however a further benefit specific of $V \rightarrow Pe^{+}e^{-}$ 
events: these processes can in fact be
used also to measure  {\em on an event-by-event basis} the actual 
dark photon decay path, provided
that the final state meson decays into at least a pair of charged particles. 
In this case, the position of the latter particles determines the 
{\em primary production} vertex,
while the $\gamma'$ decay position is determined, as usual, 
by the \epem\ one. 

For instance, for $J/\psi \rightarrow  \eta'\gamma'$ events, 
one can use the process 
$\eta' \rightarrow \eta\pi^{+}\pi^{-}$, where the $\pi^{+}\pi^{-}$ pair allows
determining exactly  
the collision point, at the price of reducing the total amount
of useful events by a factor $\sim$0.43~\cite{PDG}.
Since we are interested in millimeter scale decay paths, the photon conversion
background should not be a problem anymore. However there is a
physical background to be kept now into consideration, namely 
the Dalitz decay of the J$\psi$,  
$J/\psi \rightarrow \eta'$\epem. Its branching ratio
can be estimated to be approximately 
BR($J/\psi \rightarrow \eta'$\epem) $\sim$
BR($J/\psi \rightarrow \eta'\gamma$)$\times$0.01, so that the process is 
$\sim$10$^{6}$ times more frequent than the signal, if 
$\epsilon$=10$^{-4}$. 
However, in this case, the \epem\ and $\pi^{+}\pi^{-}$ vertices must coincide 
within the detector's resolution, $\sigma_{res}$.  Therefore a reduction of
the background by a factor 10$^{4}$ can be achieved by accepting only 
events with measured decay paths larger than 4$\sigma_{res}$. 
Moreover, background events are 
expected to have a broad \epem\ invariant mass distribution, while 
$\gamma'$ decays are a narrow resonance in such channel. Not considering
form factor effects, the number of background events, N$_{B}$, in a window of 
$\delta$m around m$_{e^+e^-}$=m$_{\gamma'}$ is given approximately 
by~\cite{RWa}

\begin{equation}
N_{B} = N_{J/\psi \rightarrow \eta'e^+e^-}\times B(\eta'\rightarrow\eta\pi^+\pi^-)
\times\frac{\delta m}{m_{\gamma'}}
  \times\frac{1}{ln((m_{J/\psi}-m_{\eta'})/2m_{e}))} 
\label{eq:eqnb}
\end{equation}

Note that the 1/m$_{\gamma'}$ dependence of equation~\ref{eq:eqnb} favours the 
observation of higher mass dark photons.  On the other hand, since the decay
path scales as 1/m$^{2}_{\gamma'}$,  vertex resolution effects point in the 
opposite direction. 

In a given experiment, therefore, two parameters should ideally be kept 
as small as
possible, $\sigma_{res}$ and $\delta$m.
Take, for instance, $\epsilon$=10$^{-4}$, 
m$_{\gamma'}$=50 MeV. According to eq.~\ref{eq:eqnb}, the number of background
events in the interesting mass bin would in this case 
be $\sim$2$\times$10$^{5}$ for $\delta_{m}$=1 MeV.  By applying the
 4$\sigma_{res}$ cut discussed above, this number reduces to $\sim$20. 
Therefore,  for $\sigma_{res}$=1(0.5) mm, the signal significance (i.e. 
the number of signal events divided by the square root of the background) 
would
be $\sim$4(6). It is important to underline that in this case the dimensions
of the beam pipe are to first order irrelevant, since we are dealing with
relatively short decay lengths. On the other hand, it has also to be 
noted, that we are here assuming full detection efficiency which might reveal
an overoptimistic assumption. As for the continuum events, only detailed 
studies based on realistic detector parameters can finally assess the
 potentials of the method. 

\section{Conclusions}
Experimental searches for a new, light, neutral boson, the ``dark photon'' or
$\gamma'$, are being pursued in many laboratories in the world, using
different detection techniques. If the $\gamma'$ is light enough and if its
couplings with SM particles are suppressed by a factor $\leq$10$^{-3}$ with
respect to those of the ordinary photon, it can acquire a relatively 
long lifetime. This fact can be exploited at \epem\ colliders by searching 
a $\gamma' \rightarrow e^{+}e^{-}$  decay vertex well separated by the
primary production one, in $e^{+}e^{-} \rightarrow \gamma'\gamma$ 
events.
The paper shows that new generation \epem\ colliders have the potentials 
to fully exploit this technique, and can explore effective couplings
down to few times 10$^{-5}$ and $\gamma'$ masses in the range 
10-100 MeV approximately. In general higher energy machines are favoured, 
because the higher Lorentz boost of the produced (light) bosons allows
a better separation of the secondary vertices. 
It is also seen, however, that the implementation of this method to 
real facilities requires a proper design of the interaction region and 
a wise choice of the tracking detector. To our knowledge, unfortunately, 
the beam pipe dimensions of SuperBelle are far from being optimal under 
this respect. On the other hand,  
since all  of the future high luminosity $\tau$-charm factories are
still in a preliminary design study phase, it is conceivable that in 
this case the interaction region can be designed such as to maximise the 
sensitivity for the proposed $\gamma'$ search technique.
Such a machine can also provide a complementary search method by the 
observation of displaced \epem\ vertices in fully reconstructed 
$J/\psi \rightarrow \eta'\gamma'$
events, therefore enhancing the interest for the construction of
such a facility. 

\section*{Acnowledgements}
I would like to thank S. Andreas and D.~Babusci for help and useful
 discussions.

\end{document}